\documentclass[prd,
               twocolumn,
			   nofootinbib,
			   superscriptadress,
			   longbibliography,
			   english]{revtex4-1}

\usepackage{slashed}
\usepackage[utf8]{inputenc}
\usepackage{babel}
\usepackage{graphics}
\usepackage{graphicx}

\usepackage{epsfig}
\usepackage{float}
\usepackage{amsmath}
\usepackage{amssymb}
\usepackage{enumerate}

\usepackage{hyperref}

\newcommand{\be}{\begin{equation}}
\newcommand{\ee}{\end{equation}}

\begin{document}

\title{Massive Schwinger model at finite $\theta$}

\author{Vicente Azcoiti}
\affiliation{Departamento de Física Teórica, Facultad de Ciencias,
  Universidad de Zaragoza \\ C/Pedro Cerbuna 12, E-50009, Zaragoza
  (Spain)}
\author{Giuseppe \surname{Di Carlo}}
\affiliation{INFN, Laboratori Nazionali del Gran Sasso, \\
I-67100 Assergi, L'Aquila (Italy)}
\author{Eduardo Follana}
\affiliation{Departamento de Física Teórica, Facultad de Ciencias,
  Universidad de Zaragoza \\ C/Pedro Cerbuna 12, E-50009, Zaragoza
  (Spain)}
\author{Eduardo Royo-Amondarain}
\email{Corresponding author: eduroyo@unizar.es}
\affiliation{Departamento de Física Teórica, Facultad de Ciencias,
  Universidad de Zaragoza \\ C/Pedro Cerbuna 12, E-50009, Zaragoza 
  (Spain)}
\author{Alejandro Vaquero Avil\'es-Casco}
\affiliation{Department of Physics and Astronomy, University of Utah, Salt Lake City, UT 84112, USA}

\begin{abstract}
	Using the approach developed in [V. Azcoiti, G. Di Carlo,
	A. Galante, V. Laliena, \textit{Phys. Lett.} \textbf{B563}, (2003)
	117], we are able to reconstruct the behavior of the massive
	1-flavor Schwinger model with a $\theta$ term and a quantized
	topological charge. We calculate the full dependence of the order
	parameter with $\theta$. Our results at $\theta = \pi$ are
	compatible with Coleman's conjecture on the phase diagram of this
	model.
\end{abstract}

\maketitle

\section{Introduction}
\label{Introduction}

The origin of dark matter is on the basis of the aim to elucidate the
existence of new low-mass, weakly interacting particles from a
theoretical, phenomenological and experimental point of view. The
light particle that has gathered the most attention is the axion,
predicted by Weinberg and Wilczek \cite{Wein78}, and Wilczek
\cite{Wilc78} in the Peccei and Quinn mechanism \cite{pq} to explain
the absence of parity and temporal invariance violations induced by
the QCD vacuum.  The axion is one of the more interesting candidates
to make the dark matter of the universe, and the axion potential, that
determines the dynamics of the axion field, plays a fundamental role
in this context.

The $QCD$ axion model relates the topological susceptibility $\chi_T$
with the axion mass $m_a$ and decay constant $f_a$ through the
relation $\chi_T = m^2_a f^2_a$. The axion mass is, on the other hand,
an essential ingredient in the calculation of the axion abundance in
the Universe. Therefore, a precise computation of the topological
properties of $QCD$ and of their temperature dependence becomes of
primordial interest in this context. Understanding the role of the
$\theta$ parameter in QCD and its connection with the strong CP
problem is one of the major challenges for high energy theorists
\cite{peccei}.


The calculation of the topological susceptibility in QCD is already a
challenge, but calculating the complete potential requires a strategy
to deal with the so called sign problem, that is, the presence of a
highly oscillating term in the path integral.  In fact euclidean
lattice gauge theory, our main non-perturbative tool for studying QCD
from first principles, has not been able to help us much because of
the imaginary contribution to the action coming from the
$\theta$ term, that prevents the applicability of the importance
sampling method \cite{vicari}. This is the main reason why the only
progress in the analysis of the finite temperature $\theta$ dependence
of the vacuum energy density in pure gauge QCD, outside of
approximations, reduces to the computation of the first few
coefficients in the expansion of the free energy density in powers of
$\theta$ \cite{bdpv}, and the situation in full QCD with dynamical
fermions is, on the other hand, even worse
\cite{martinelli,petre,javier,vicente,Wolfgang,vicente2}.

Much experience has been developed in the last years by our group in
this field, both in the elaboration of efficient algorithms to
simulate systems with a theta-vacuum term overcoming the severe sign
problem \cite{m1,Azco03}, as well as in the application of these
approaches to the computation of the vacuum energy density and
topological charge density for several interesting physical systems
\cite{adgl,adg,afv,adfg,acfg}. Our purpose is to take advantage of
this experience to apply these approaches to the computation of the
$\theta$ dependence of the QCD vacuum energy density.

As a first step in this ambitious program we present in this paper an
analysis on the $\theta$ dependence of a toy model for QCD, the
Schwinger model, on the lattice.  Strictly speaking, the Schwinger
model in the continuum is not asymptotically free, as QCD, since it is
super-renormalizable and the Callan-Symanzik $\beta$-function
vanishes. However, in the lattice version, since the continuum
coupling is dimensionful, the continuum theory is reached at infinite
inverse square gauge coupling $\beta = 1/e^2a^2$, much in the same way
as four-dimensional asymptotically free gauge theories such as
QCD. Furthermore the model is confining \cite{kogut1}, exactly
solvable at zero fermion mass, has non-trivial topology and shows
explicitly the $U_A(1)$ axial anomaly \cite{kogut2} through a
non-vanishing value of the chiral condensate in the chiral limit in
the one-flavor case.  These are basically the reasons why this model
has been extensively used as a toy model for $QCD$.

For two dimensional systems such as the Schwinger model with a
$\theta$ term, there exist numerical methods such as the Hamiltonian
method \cite{kogut3,bis} and the Grassmann tensor renormalization
group method \cite{shimizu} that have been applied
successfully. However, such methods are currently only applicable to
two-dimensional systems, whereas our aim is to test a method that
should, in principle, be applicable also to four-dimensional theories
such as QCD.

The paper is organized as follows; in Section \ref{continuum} we
summarize some relevant features of the Schwinger model with
topological term. Since our second proposal to analyze physical
systems with a topological term in the action \cite{Azco03} has been found
to be particularly well suited to bypass the sign problem in
asymptotically free gauge theories, we decided to apply it, and
Sec. \ref{methodII} contains a brief review of the main steps of
the method. In Sec. \ref{lattice} we give details on the lattice
setup and the computer simulations. Sec. \ref{results} shows our
results for the topological charge density as a function of $\theta$
at several fermion masses and gauge couplings, and finally in Sec.
\ref{conclusions} we report our conclusions.

\section{The massive Schwinger model with a $\theta$ term}
\label{continuum}

The Schwinger model is Quantum Electrodynamics in
1+1-dimensions\cite{Schwinger1962}. The euclidean continuum action reads

\begin{eqnarray}
S = \int d^2x \{\bar\psi(x)\gamma_\mu\left(\partial_\mu 
       + ieA_\mu(x)\right)\psi(x) \nonumber\\
  + m \bar\psi(x)\psi(x) + {1\over4} F^2_{\mu\nu}(x)\},
\label{uno}
\end{eqnarray}

\noindent
where $m$ is the fermion mass and $e$ is the electric charge or gauge
coupling, which has the same dimension as $m$. After a simple
rescaling of the fields the action can be written as

\begin{eqnarray}
S = \int d^2x \{\bar\psi(x)\gamma_\mu\left(\partial_\mu 
       + iA_\mu(x)\right)\psi(x) \nonumber\\
        + m \bar\psi(x)\psi(x) + {1\over{4e^2}} F^2_{\mu\nu}(x)\},
\label{dos}
\end{eqnarray}

\noindent
where $F_{\mu\nu}(x) = \partial_\mu A_\nu(x) - \partial_\nu A_\mu(x)$ and $\gamma_\mu$ are 
$2\times 2$ matrices satisfying the algebra

\begin{equation}
\{\gamma_\mu, \gamma_\nu\} = 2 g_{\mu\nu}.
\end{equation}

At the classical level this action is invariant in the chiral limit
under the $U_A(1)$ global transformations

\begin{eqnarray}
\psi &\rightarrow & e^{i\alpha\gamma_5}\psi, \\
\bar\psi &\rightarrow & \bar\psi e^{i\alpha\gamma_5},
\end{eqnarray}

\noindent
leading to the conservation of the axial current

\begin{equation}
J^A_\mu(x) = \bar\psi(x)\gamma_\mu\gamma_5\psi(x).
\label{axialcurrent}
\end{equation}

\noindent
However the axial symmetry is broken at the quantum level because of
the axial anomaly. The divergence of the axial current is

\begin{equation}
\partial_\mu J^A_\mu(x) = {1\over{2\pi}} \epsilon_{\mu\nu} F_{\mu\nu}(x),
\label{currentdivergence}
\end{equation}

\noindent
with $\epsilon_{\mu\nu}$ the antisymmetric tensor, and therefore does
not vanish.  The axial anomaly induces a topological $\theta$ term in
the action of the form

\begin{equation}
S_{top} = {{i\theta}\over{4\pi}} \int d^2x \epsilon_{\mu\nu} F_{\mu\nu}(x),
\label{topologicalaction}
\end{equation}

\noindent
where the topological charge $Q = {1\over{4\pi}} \int d^2x
\epsilon_{\mu\nu} F_{\mu\nu}(x)$ is an integer.

The purpose of this paper is to analyze the $\theta$ dependence of the
model described by the action (\ref{dos})+(\ref{topologicalaction})

\begin{eqnarray}
S = \int d^2x \{\bar\psi\gamma_\mu\left(\partial_\mu + iA_\mu\right)\psi +
m \bar\psi\psi \nonumber\\
+ {1\over{4e^2}} F^2_{\mu\nu}
+ {{i\theta}\over{4\pi}}\epsilon_{\mu\nu} F_{\mu\nu}\}.
\label{action}
\end{eqnarray}

A simple analysis of this model on the lattice suggests that it should
undergo a phase transition at some intermediate fermion mass $m$ and
$\theta=\pi$, even at finite lattice spacing.  Indeed the lattice
model is analytically solvable in the infinite fermion mass limit
(pure gauge two-dimensional electrodynamics with topological term)
\cite{puregauge}, and it is well known that the density of topological
charge approaches a non-vanishing vacuum expectation value at
$\theta=\pi$ for any value of the inverse square gauge coupling
$\beta$, exhibiting spontaneous symmetry breaking. On the other hand
by expanding the vacuum energy density in powers of $m$, treating the
fermion mass as a perturbation \cite{smilga}, one gets for the vacuum
expectation value of the density of topological charge the following
$\theta$ dependence:

\begin{equation}
\langle -i q\rangle = m\Sigma sin\theta + \frac{1}{2} m^2 \sin \left(2\theta\right) 
\left( \chi_P - \chi_S\right) + \cdots,
\label{qchiral}
\end{equation}

\noindent
with $\Sigma$ the vacuum expectation value of the chiral condensate in
the chiral limit and at $\theta=0$ ($\Sigma =
e^{\gamma_e}e/2\pi^{3/2}$ in the continuum limit), and $\chi_P$ and
$\chi_S$ the pseudoscalar and scalar susceptibilities
respectively. Equation (\ref{qchiral}) shows how the $Z_2$ symmetry at
$\theta=\pi$ is realized order by order in the perturbative expansion
of the topological charge in powers of the fermion mass $m$, and
therefore a critical point separating the large and small fermion mass
phases is expected.

Indeed the model was analyzed in the continuum by Coleman in
\cite{coleman}, where he conjectured the existence of a phase
transition at $\theta=\pi$, and some intermediate fermion mass $m$
separating a ''weak coupling'' phase (${e\over m}<<1$), where the
$Z_2$ symmetry of the model at $\theta=\pi$ is spontaneously broken,
from a ''strong coupling'' phase (${e\over m}>>1$) where the $Z_2$
symmetry is realized in the vacuum. This conjecture was corroborated
in \cite{kogut3,bis} using the lattice Hamiltonian approach with
staggered fermions, and more recently in \cite{shimizu} using the
Grassmann tensor renormalization group and Wilson fermions.

\section{Computing the order parameter as a function of $\theta$}
\label{methodII}

To compute the $\theta$ dependence of the density of topological
charge we use the approach proposed in reference \cite{Azco03}. The only
assumption in this approach is the absence of phase transitions at
real values of $\theta$ except at most at $\theta=\pi$. The method is
based in extrapolating a suitably defined function to the origin. This
function turns out to be very smooth in all the cases considered up to
now \cite{adgl,adg,afv,adfg},
and this makes us confident on the whole procedure. Here we summarize
the main steps.

From numerical simulations of our physical system at imaginary values
of $\theta= -ih$ (real values of $h$), which are free from the severe
sign problem, we compute the density of topological charge $q(-ih)$ as
a function of $h$, and introduce the following functions:

\begin{eqnarray}
z = \cosh \frac{h}{2}, \\
y(z) = \frac {q(-ih)}{\tanh \frac{h}{2}}.
\label{ypsilon}
\end{eqnarray}

The procedure to find out the density of topological charge at real
values of $\theta$ relies on scaling transformations \cite{Azco03}. We
define the function $y_\lambda\left(z\right)$ as
\begin{equation}
y_\lambda\left(z\right) = y\left(e^{\frac{\lambda}{2}}z\right).
\label{Is-29}
\end{equation}
For negative values of $\lambda$, the function
$y_\lambda\left(z\right)$ allows us to calculate the order parameter
$\left(\tanh\frac{h}{2}\,y\left(z\right)\right)$ below the threshold
$z = 1$. If $y\left(z\right)$ is non-vanishing for any positive
$z$,\footnote{Even though the possibility of a vanishing
  $y\left(z\right)$ for some value $z>0$ cannot be
  completely excluded, it does not happen for any of the analytically solvable
  models we know.} then we can plot $y_\lambda/y$ against
$y$. Furthermore, in the case that $y_\lambda/y$ is a smooth function
of $y$ close to the origin, then we can rely on a simple extrapolation
to $y=0$. Of course, a smooth behavior of $y_\lambda/y$ cannot be
taken for granted; however no violations of this rule have been found
in the exactly solvable models.

The behavior of the model at $\theta = \pi$ can be ascertained from
this extrapolation. At $\theta = \pi$ the model has the same $Z_2$
symmetry than at $\theta = 0$. We can define an effective exponent
$\gamma_\lambda$ by
\begin{equation}
\gamma_\lambda = \frac{2}{\lambda}\ln\left(\frac{y_\lambda}{y}\right).
\label{Is-30}
\end{equation}
As $z\rightarrow0$, the order parameter
$\tan\frac{\theta}{2}\,y\left(\cos\frac{\theta}{2}\right)$ behaves as
$\left(\pi-\theta\right)^{\gamma_\lambda-1}$. Therefore, a value of
$\gamma_\lambda = 1$ implies spontaneous symmetry breaking at $\theta
= \pi$. A value between $1 < \gamma_\lambda < 2$ signals a second
order phase transition, and the corresponding susceptibility
diverges. Finally, if $\gamma_\lambda = 2$, the symmetry is realized
(at least for the selected order parameter), there is no phase
transition and the free energy is analytic at
$\theta=\pi$.\footnote{Other possibilities are allowed, for instance,
  any $\gamma_\lambda > 1,\quad\gamma_\lambda\in\mathbb{N}$ leads to
  symmetry realization for the order parameter at $\theta=\pi$ and to
  an analytic free energy. If $\gamma_\lambda$ lies between two
  natural numbers, $p < \gamma_\lambda < q,\quad p,q\in\mathbb{N}$,
  then a transition of order $q$ takes place.}

\begin{figure}[h!]
  \centerline{\includegraphics[angle=90,width=\linewidth]
    {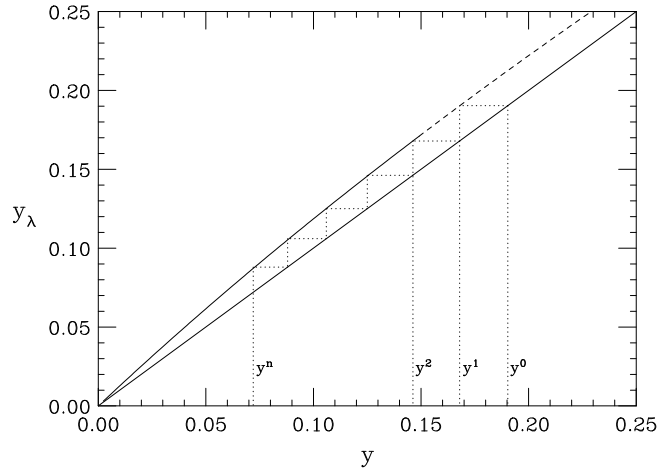}}
\caption{\label{reconstruction}Iterative method used to compute the
  different values of $y(z)$. $y_\lambda$ is plotted as a function of
  $y$ using a dashed line in the region where direct measurements are
  available, and a continuous line in the extrapolated region. The
  straight continuous line represents $y_\lambda=y$.}
\end{figure}

We can take the information contained in the quotient
$\frac{y_\lambda}{y}\left(y\right)$, and calculate the order parameter
for any value of $\theta$ through an iterative procedure
\cite{Azco03}. The outline of the procedure is the following:
\begin{enumerate}[i.]
\item Beginning from a point $y\left(z_i\right) = y_i$, we find the
  value $y_{i+1}$ such that $y_\lambda=y_i$. By definition,
  $y_{i+1}=y\left(e^{\frac{-\lambda}{2}}z_i\right)$.
\item Replace $y_i$ by $y_{i+1}$, and start again.
\end{enumerate}
The procedure is repeated until enough values of $y$ are know for
$z<1$ (see Fig. \ref{reconstruction}). This method can be used for any
model, as long as our assumptions of smoothness and absence of
singular behavior are verified during the numerical
computations. The reliability of our approach in practical
applications is better when the following conditions are met:
\begin{enumerate}[a.]
\item $y(z)$ takes small values for values of $z$ of order 1.
\item The dependence on $y$ of the functions $y_\lambda/y$ and $\gamma_\lambda$
is soft enough to allow a reliable extrapolation.
\end{enumerate}
In the one-dimensional Ising model within an imaginary magnetic field
these two properties are realized in the low temperature regime
\cite{Azco03}, and the two and three-dimensional models also show a very
good behavior in this regime \cite{afv}. Indeed the relevant feature,
at least in what concerns point $a$, is that, at low temperatures, the
magnetic susceptibility at small values of the real external magnetic
field takes small values. In the more interesting case of
asymptotically free models, the analogue of the magnetic
susceptibility is the topological susceptibility, and it is well known
that topological structures are strongly suppressed near the continuum
limit. Therefore, and on qualitative grounds, we expect a good
implementation of our method in the Schwinger model or in $QCD$, at
least close enough to the continuum limit. In fact, concerning
asymptotically free models, the method was successfully applied to the
analysis of the continuum $\theta-$dependence of $CP^9$ \cite{adgl},
showing a very good realization of conditions $a$ and $b$. In the more
general cases we should find out whether the model agrees with these
two conditions or not, and pleasant surprises are not excluded
\cite{adg}, \cite{adfg}.

\section{Details of the simulation}
\label{lattice}

We use the lattice version of the continuum action (\ref{action}) with
staggered fermions and standard Wilson form for the pure gauge
part. It reads as follows,

\begin{eqnarray}
S &=& {1\over2} \sum_{n, \mu}
\eta_\mu(n)\bar\chi(n)\{U_\mu(n)\chi(n+\hat{\mu}) \nonumber\\
& &-U^\dagger_\mu(n-\hat{\mu})\chi(n-\hat{\mu})\} + m \sum_n \bar\chi(n)\chi(n) \nonumber\\
& &-\beta \sum_n Re
\left(U_1(n)U_2(n+\hat{1})U^\dagger_1(n+\hat{2})U^\dagger_2(n)\right) \nonumber\\
& &- i\theta \sum_n q(n),
\label{latticeaction}
\end{eqnarray}
where the notation is standard. The compact gauge variable $U_\mu(n)$
is related to the non-compact gauge field $A_\mu(n)$ in the usual way
\begin{equation}
U_\mu(n) = e^{ia A_\mu(n)} = e^{i \phi_\mu(n)},
\end{equation}
\noindent
with $a$ the lattice spacing, and the local topological charge $q(n)$
is given by
\begin{eqnarray}
q(n) &=& {1\over{2\pi}} \left(\phi_1(n) + \phi_2(n+\hat{1}) \right.\nonumber\\ 
& &- \left.\phi_1(n+\hat{2}) - \phi_2(n)\hskip 0,2cm mod \hskip 0,2cm 2\pi\right).
\label{localtopcharge}
\end{eqnarray}
An important point is that this charge is quantized, and therefore the
partition function of the model has exact $2\pi$ periodicity in
$\theta$, as is the case in the continuum theory.\footnote{This is an
  important difference with the approach in \cite{Gattringer}, which
  makes a comparison with our results at finite lattice spacing
  difficult.}

We will analyze in what follows the model given by action
(\ref{latticeaction}), taking the square root of the fermion
determinant in order to describe only one flavor. There is ample
evidence that this procedure leads to the correct physics in the
continuum limit, including the effects of the anomaly.  For example,
the Microcanonical Fermion Average (MFA) approach \cite{mfa} was
applied years ago to simulate the one-flavor Schwinger model on the
lattice at $\theta=0$, using the standard Wilson action for the gauge
field and staggered fermions. The results \cite{sch_noi} reproduce the
exact value of the chiral condensate in the chiral continuum limit up
to 3 decimal places.\footnote{See also \cite{Durr1,Durr2}.}

As explained in Sec. \ref{methodII}, in order to find the dependence
on $\theta$ of the density of topological charge $q$, we need to
compute the expected value of $q$ for imaginary values of $\theta$. In
this case standard Monte Carlo algorithms work well, and we can sample
the distribution $e^{-S}$ generated by (\ref{latticeaction}) with any
of these methods. We have used a standard Metropolis approach, trying
to update each link sequentially in every sweep.

We want to use an exact Monte Carlo method, and the fermionic part of
the action forces us to recompute the whole fermion determinant at
each attempt to update one link, that is, $N$ times each
sweep. Indeed, this is the most expensive part of the algorithm. All
our lattices are of size $16\times16$, and we computed the eigenvalues
of the fermion matrix with the GNU Scientific Library, taking
advantage of the standard even-odd decomposition of the staggered
fermions Dirac operator. The simulations have been run at the U-LITE
computer facility at the INFN National Laboratories of Gran Sasso.

We present in the following section results for several masses $m$,
gauge couplings $\beta$, and fields $h$ (imaginary $\theta$). At each
point of the parameter space, we run the algorithm and take up to 100k
measurements, each one made every ten sweeps. Between 5-10\% of the
initial configurations are discarded for thermalization. The errors of
the expected values are estimated by a standard jackknife binning.

\section{Results}
\label{results}

\begin{figure}[h!]
  \centerline{\includegraphics[angle=0,width=\linewidth]
    {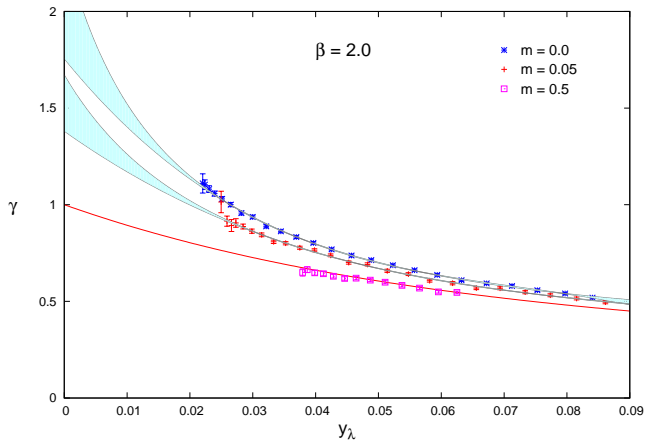}}
\caption{\label{beta2_op}Exponent $\gamma$ for $\beta = 2.0$ and
  various fermion masses. The shaded areas give an estimation of the
  ambiguity in the extrapolation to $y_\lambda = 0$.}
\end{figure}

\begin{figure}[h!]
  \centerline{\includegraphics[angle=0,width=\linewidth]
    {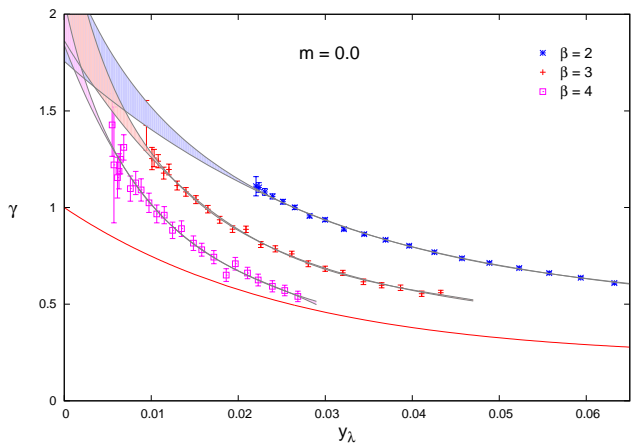}}
\caption{\label{m0_op}Exponent $\gamma$ for $m = 0.0$ and various
  coupling constants. The shaded areas give an estimation of the
  ambiguity in the extrapolation to $y_\lambda = 0$.}
\end{figure}

Our results for the exponent $\gamma$ are summarized in Figures
\ref{beta2_op} and \ref{m0_op}. As is apparent in Fig.
\ref{beta2_op}, the behavior at fixed $\beta$ is very different as we
vary the fermion mass. The data corresponding to $m = 0.5$ lie
essentially on top of the analytic $m = \infty$ curve (that is, pure
gauge theory \cite{puregauge}), and therefore this mass corresponds to
the phase with broken symmetry at $\theta = \pi$. On the other hand,
the data for both the $m = 0.05$ and the $m = 0$ case extrapolate to a
value clearly above 1, indicating symmetry restoration at $\theta =
\pi$, although our data are not precise enough to make a definite
statement on the value of $\gamma$. In Fig. \ref{m0_op} we present
the results at fixed $m = 0$ for the various coupling constants we
have studied. We see as before a clear extrapolation to a value of
$\gamma$ above 1,\footnote{This is also the case for $\beta = 3$, $m =
  0.05$, which is not shown in these figures.} in stark contrast with
the pure gauge theory case.

\begin{figure}[h!]
  \centerline{\includegraphics[angle=0,width=\linewidth]
    {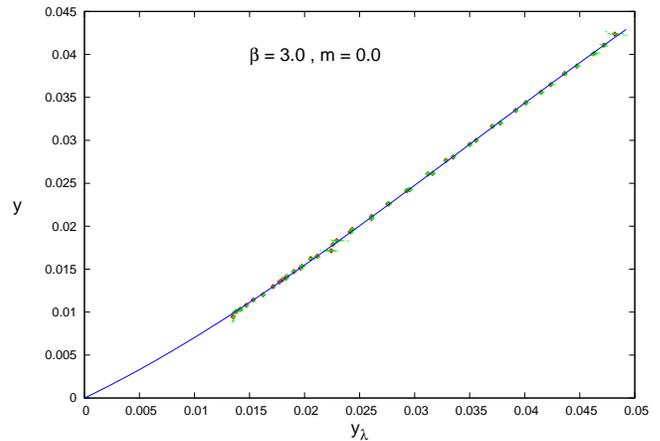}}
\caption{\label{y_lambda}Fit of $y$ versus $y_\lambda$.}
\end{figure}

We have also been able to extract the full dependence of the order
parameter as a function of the angle, $q(\theta)$. In Fig.
\ref{y_lambda} we show details of the fit $y$ versus $y_\lambda$ for a
particular value of the parameters, in order to give an idea of the
precision of our data. This will be followed by the iterative
procedure depicted in Sec. \ref{methodII} in order to produce the
curve $q(\theta)$. Regarding the error estimation, it is impossible to
follow the error propagation from the values of $q$ at imaginary
$\theta$ until the final result of $q(\theta)$.  To overcome this
impasse we proceed in the following way: first we generate 20 sets of
fake data for $q(h)$ having the same mean value and distribution of
the actual Monte Carlo data; then we compute, following the same
scheme (fit of $y$ vs. $y_\lambda$ and iterative procedure), 20
realization of $q(\theta)$; from the distribution of these values
around the curve computed from the real data, we can infer the error
associated to each value of $q(\theta)$, which will be shown in the
following figures as a shaded band.

\begin{figure}[h!]
  \centerline{\includegraphics[angle=0,width=\linewidth]
    {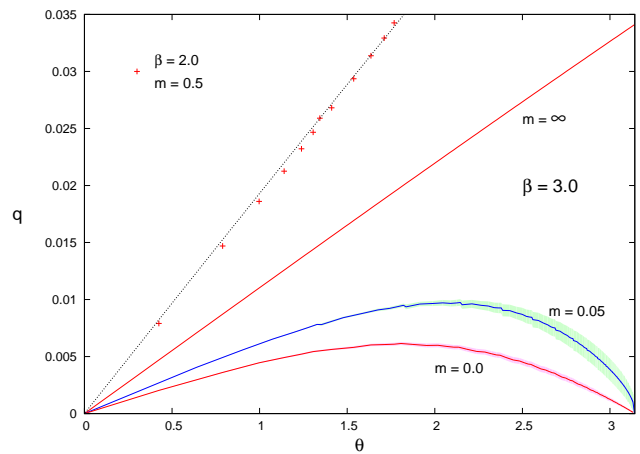}}
\caption{\label{beta3_q}Order parameter as a function of $\theta$. The
  data at $m = 0.0$ and $m = 0.05$ correspond to $\beta = 3.0$,
  whereas the points at $m = 0.5$ correspond to $\beta = 2$. The
  continuous line labeled $m = \infty$ is the pure gauge analytic
  result for $\beta = 3.0$, whereas the dotted line is the
  corresponding analytic result for $\beta = 2.0$. }
\end{figure}

\begin{figure}[h!]
  \centerline{\includegraphics[angle=0,width=\linewidth]
    {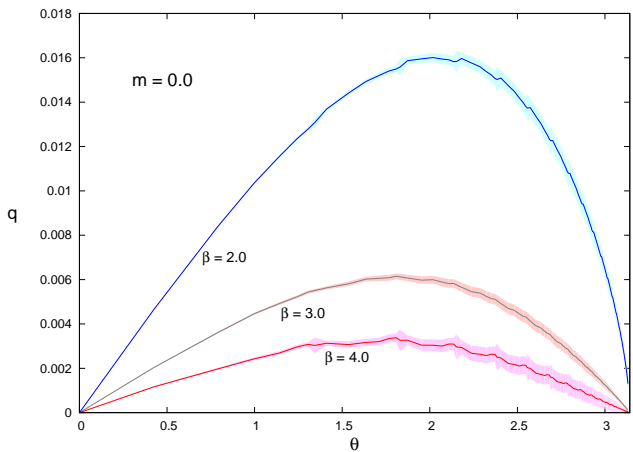}}
\caption{\label{m0_q}Order parameter as a function of $\theta$, at
  $m = 0.0$ and different coupling constants.}
\end{figure}

\begin{figure}[h!]
  \centerline{\includegraphics[angle=0,width=\linewidth]
    {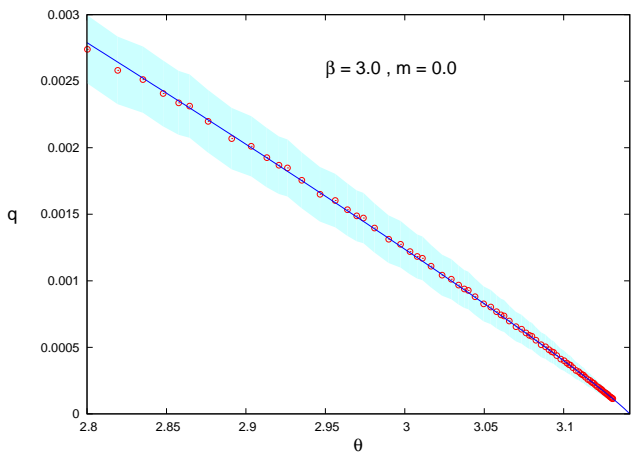}}
\caption{\label{beta3_m0_q}Order parameter as a function of $\theta$
  near $\theta = \pi$.}
\end{figure}

\begin{table}[h!]
\begin{center}
  \caption{\label{epsilon} }
  \begin{tabular}{ccc}
 $\beta$ &   $m$ & $\epsilon$ \\
\hline
2.0 & 0.0  & 0.67(4) \\
2.0 & 0.05 & 0.43(5) \\
3.0 & 0.0  & 0.92(7) \\
3.0 & 0.05 & 0.70(21) \\
4.0 & 0.0  & 0.94(19)
  \end{tabular}
\end{center}
\end{table}

In Fig. \ref{beta3_q} we present $q(\theta)$ at $\beta = 3$ for two
masses in the symmetry restored phase, as well as at $\beta = 2$ and
$m = 0.5$, in the symmetry broken phase (and also the corresponding
analytic results for the pure gauge case at both values of $\beta$ for
comparison).

In Fig. \ref{m0_q} we show the results for $m = 0$ and the three
different values of the coupling constant we have simulated. We can
clearly see the restoration of the symmetry as we approach $\theta =
\pi$. In Fig. \ref{beta3_m0_q} we show, for $\beta = 3.0$ and $m =
0$, the order parameter $q(\theta)$ in the vicinity of $\theta =
\pi$. Fitting $q(\theta)$ near $\theta = \pi$ in the symmetry restored
phase allows us to extract the exponent $(\pi - \theta)^ \epsilon$,
which is related to $\gamma$ by $\epsilon = \gamma - 1$.\footnote{The
  numerical procedure used to extract the two exponents is different,
  and therefore the results, although compatible within errors, will
  also be different.} We present in Table \ref{epsilon} our results
for $\epsilon$.

\begin{figure}[h!]
  \centerline{\includegraphics[angle=0,width=\linewidth]
    {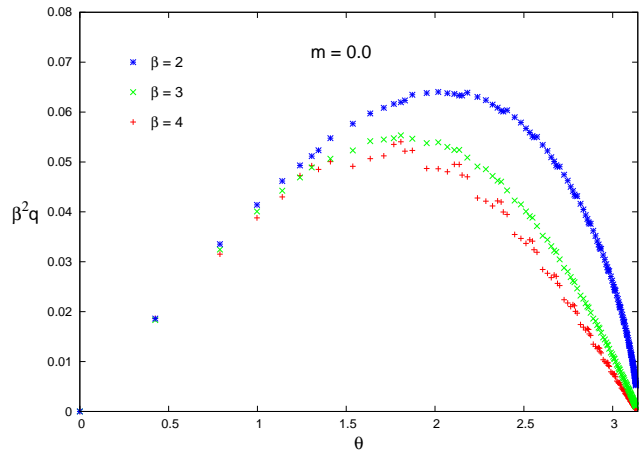}}
\caption{\label{scaling}Rescaled topological charge density at $m=0.0$
  and different coupling constants.}
\end{figure}

To finish this Sec. we want to discuss a little bit more on the
results for the massless Schwinger model reported in Fig.
\ref{m0_q}. It is well known that the continuum formulation of the
massless Schwinger model shows no $\theta$ dependence, because the
$\theta$ term in the action can be canceled by an anomalous chiral
transformation which does not change the fermion-gauge action if the
fermion mass vanishes. Hence the non-trivial $\theta$ dependence of
the density of topological charge shown in Fig. \ref{m0_q} may seem
surprising. However, the massless staggered Dirac operator does not
have exact zero-modes, and therefore, for a given gauge configuration,
a nonzero value of the quantized topological charge $Q$ does not
imply the existence of a corresponding number of zero-modes in the
staggered Dirac operator, as would be the case, for example, with the
overlap Dirac operator.  What we should expect instead is that, as we
approach the continuum limit, the topological charge density
vanishes. This is indeed what seems to happen, as is suggested by the
results of Fig. \ref{scaling}.

\section{Conclusions and outlook}
\label{conclusions}

All our results are compatible with the standard lore on this model,
and in particular with Coleman's conjecture on the existence of two
distinct phases at $\theta = \pi$, a symmetry breaking phase at large
mass, and a symmetry restored phase at small mass.

Our simulations are a proof of concept, and are not extensive enough
to determine precisely the position of the critical mass at $\theta =
\pi$ or its properties in detail. But the important point is that we
have succeeded in calculating the full dependence of the order
parameter in $\theta$ in a gauge theory with fermions and a quantized
topological charge, using a method that should, in principle, work
also in higher dimensional theories.

The aim of this calculation was to test the method developed in
\cite{Azco03} in a fermionic gauge theory, as a first step towards
applying it to QCD with a $\theta$ term. In light of the excellent
results obtained, we expect the method to be applicable also in this
case.

\section{Acknowledgments}
This work was funded by Ministerio de Economía y Competitividad/Fondo
Europeo de Desarrollo Regional Grants No. FPA2015-65745-P and Diputación
General de Aragón-Fondo Social Europeo Grant No. 2015-E24/2.


\begin{thebibliography}{33}%
\makeatletter
\providecommand \@ifxundefined [1]{%
 \@ifx{#1\undefined}
}%
\providecommand \@ifnum [1]{%
 \ifnum #1\expandafter \@firstoftwo
 \else \expandafter \@secondoftwo
 \fi
}%
\providecommand \@ifx [1]{%
 \ifx #1\expandafter \@firstoftwo
 \else \expandafter \@secondoftwo
 \fi
}%
\providecommand \natexlab [1]{#1}%
\providecommand \enquote  [1]{``#1''}%
\providecommand \bibnamefont  [1]{#1}%
\providecommand \bibfnamefont [1]{#1}%
\providecommand \citenamefont [1]{#1}%
\providecommand \href@noop [0]{\@secondoftwo}%
\providecommand \href [0]{\begingroup \@sanitize@url \@href}%
\providecommand \@href[1]{\@@startlink{#1}\@@href}%
\providecommand \@@href[1]{\endgroup#1\@@endlink}%
\providecommand \@sanitize@url [0]{\catcode `\\12\catcode `\$12\catcode
  `\&12\catcode `\#12\catcode `\^12\catcode `\_12\catcode `\%12\relax}%
\providecommand \@@startlink[1]{}%
\providecommand \@@endlink[0]{}%
\providecommand \url  [0]{\begingroup\@sanitize@url \@url }%
\providecommand \@url [1]{\endgroup\@href {#1}{\urlprefix }}%
\providecommand \urlprefix  [0]{URL }%
\providecommand \Eprint [0]{\href }%
\providecommand \doibase [0]{http://dx.doi.org/}%
\providecommand \selectlanguage [0]{\@gobble}%
\providecommand \bibinfo  [0]{\@secondoftwo}%
\providecommand \bibfield  [0]{\@secondoftwo}%
\providecommand \translation [1]{[#1]}%
\providecommand \BibitemOpen [0]{}%
\providecommand \bibitemStop [0]{}%
\providecommand \bibitemNoStop [0]{.\EOS\space}%
\providecommand \EOS [0]{\spacefactor3000\relax}%
\providecommand \BibitemShut  [1]{\csname bibitem#1\endcsname}%
\let\auto@bib@innerbib\@empty
\bibitem [{\citenamefont {Weinberg}(1978)}]{Wein78}%
  \BibitemOpen
  \bibfield  {author} {\bibinfo {author} {\bibfnamefont {Steven}\ \bibnamefont
  {Weinberg}},\ }\bibfield  {title} {\enquote {\bibinfo {title} {A new light
  boson?}}\ }\href {\doibase 10.1103/PhysRevLett.40.223} {\bibfield  {journal}
  {\bibinfo  {journal} {Phys. Rev. Lett.}\ }\textbf {\bibinfo {volume} {40}},\
  \bibinfo {pages} {223--226} (\bibinfo {year} {1978})}\BibitemShut {NoStop}%
\bibitem [{\citenamefont {Wilczek}(1978)}]{Wilc78}%
  \BibitemOpen
  \bibfield  {author} {\bibinfo {author} {\bibfnamefont {F.}~\bibnamefont
  {Wilczek}},\ }\bibfield  {title} {\enquote {\bibinfo {title} {Problem of
  strong $p$ and $t$ invariance in the presence of instantons},}\ }\href
  {\doibase 10.1103/PhysRevLett.40.279} {\bibfield  {journal} {\bibinfo
  {journal} {Phys. Rev. Lett.}\ }\textbf {\bibinfo {volume} {40}},\ \bibinfo
  {pages} {279--282} (\bibinfo {year} {1978})}\BibitemShut {NoStop}%
\bibitem [{\citenamefont {Peccei}\ and\ \citenamefont {Quinn}(1977)}]{pq}%
  \BibitemOpen
  \bibfield  {author} {\bibinfo {author} {\bibfnamefont {R.~D.}\ \bibnamefont
  {Peccei}}\ and\ \bibinfo {author} {\bibfnamefont {Helen~R.}\ \bibnamefont
  {Quinn}},\ }\bibfield  {title} {\enquote {\bibinfo {title} {$\mathrm{CP}$},}\
  }\href {\doibase 10.1103/PhysRevLett.38.1440} {\bibfield  {journal} {\bibinfo
   {journal} {Phys. Rev. Lett.}\ }\textbf {\bibinfo {volume} {38}},\ \bibinfo
  {pages} {1440--1443} (\bibinfo {year} {1977})}\BibitemShut {NoStop}%
\bibitem [{\citenamefont {Peccei}(2010)}]{peccei}%
  \BibitemOpen
  \bibfield  {author} {\bibinfo {author} {\bibfnamefont {R.~D.}\ \bibnamefont
  {Peccei}},\ }\bibfield  {title} {\enquote {\bibinfo {title} {{Why PQ?}}}\
  }\bibfield  {booktitle} {\emph {\bibinfo {booktitle} {{Proceedings,
  International Conference on Axions 2010: Gainesville, Florida, January 15-17,
  2010}}},\ }\href {\doibase 10.1063/1.3489562} {\bibfield  {journal} {\bibinfo
   {journal} {AIP Conf. Proc.}\ }\textbf {\bibinfo {volume} {1274}},\ \bibinfo
  {pages} {7--13} (\bibinfo {year} {2010})},\ \Eprint
  {http://arxiv.org/abs/1005.0643} {arXiv:1005.0643 [hep-ph]} \BibitemShut
  {NoStop}%
\bibitem [{\citenamefont {Vicari}\ and\ \citenamefont
  {Panagopoulos}(2009)}]{vicari}%
  \BibitemOpen
  \bibfield  {author} {\bibinfo {author} {\bibfnamefont {Ettore}\ \bibnamefont
  {Vicari}}\ and\ \bibinfo {author} {\bibfnamefont {Haralambos}\ \bibnamefont
  {Panagopoulos}},\ }\bibfield  {title} {\enquote {\bibinfo {title} {{Theta
  dependence of SU(N) gauge theories in the presence of a topological term}},}\
  }\href {\doibase 10.1016/j.physrep.2008.10.001} {\bibfield  {journal}
  {\bibinfo  {journal} {Phys. Rept.}\ }\textbf {\bibinfo {volume} {470}},\
  \bibinfo {pages} {93--150} (\bibinfo {year} {2009})},\ \Eprint
  {http://arxiv.org/abs/0803.1593} {arXiv:0803.1593 [hep-th]} \BibitemShut
  {NoStop}%
\bibitem [{\citenamefont {Bonati}\ \emph {et~al.}(2013)\citenamefont {Bonati},
  \citenamefont {D'Elia}, \citenamefont {Panagopoulos},\ and\ \citenamefont
  {Vicari}}]{bdpv}%
  \BibitemOpen
  \bibfield  {author} {\bibinfo {author} {\bibfnamefont {Claudio}\ \bibnamefont
  {Bonati}}, \bibinfo {author} {\bibfnamefont {Massimo}\ \bibnamefont
  {D'Elia}}, \bibinfo {author} {\bibfnamefont {Haralambos}\ \bibnamefont
  {Panagopoulos}}, \ and\ \bibinfo {author} {\bibfnamefont {Ettore}\
  \bibnamefont {Vicari}},\ }\bibfield  {title} {\enquote {\bibinfo {title}
  {Change of $\ensuremath{\theta}$ dependence in 4d $\mathrm{SU}(n)$ gauge
  theories across the deconfinement transition},}\ }\href {\doibase
  10.1103/PhysRevLett.110.252003} {\bibfield  {journal} {\bibinfo  {journal}
  {Phys. Rev. Lett.}\ }\textbf {\bibinfo {volume} {110}},\ \bibinfo {pages}
  {252003} (\bibinfo {year} {2013})}\BibitemShut {NoStop}%
\bibitem [{\citenamefont {Bonati}\ \emph {et~al.}(2016)\citenamefont {Bonati},
  \citenamefont {D'Elia}, \citenamefont {Mariti}, \citenamefont {Martinelli},
  \citenamefont {Mesiti}, \citenamefont {Negro}, \citenamefont {Sanfilippo},\
  and\ \citenamefont {Villadoro}}]{martinelli}%
  \BibitemOpen
  \bibfield  {author} {\bibinfo {author} {\bibfnamefont {Claudio}\ \bibnamefont
  {Bonati}}, \bibinfo {author} {\bibfnamefont {Massimo}\ \bibnamefont
  {D'Elia}}, \bibinfo {author} {\bibfnamefont {Marco}\ \bibnamefont {Mariti}},
  \bibinfo {author} {\bibfnamefont {Guido}\ \bibnamefont {Martinelli}},
  \bibinfo {author} {\bibfnamefont {Michele}\ \bibnamefont {Mesiti}}, \bibinfo
  {author} {\bibfnamefont {Francesco}\ \bibnamefont {Negro}}, \bibinfo {author}
  {\bibfnamefont {Francesco}\ \bibnamefont {Sanfilippo}}, \ and\ \bibinfo
  {author} {\bibfnamefont {Giovanni}\ \bibnamefont {Villadoro}},\ }\bibfield
  {title} {\enquote {\bibinfo {title} {{Axion phenomenology and
  $\theta$-dependence from $N_f = 2+1$ lattice QCD}},}\ }\href {\doibase
  10.1007/JHEP03(2016)155} {\bibfield  {journal} {\bibinfo  {journal} {JHEP}\
  }\textbf {\bibinfo {volume} {03}},\ \bibinfo {pages} {155} (\bibinfo {year}
  {2016})},\ \Eprint {http://arxiv.org/abs/1512.06746} {arXiv:1512.06746
  [hep-lat]} \BibitemShut {NoStop}%
\bibitem [{\citenamefont {Petreczky}\ \emph {et~al.}(2016)\citenamefont
  {Petreczky}, \citenamefont {Schadler},\ and\ \citenamefont {Sharma}}]{petre}%
  \BibitemOpen
  \bibfield  {author} {\bibinfo {author} {\bibfnamefont {Peter}\ \bibnamefont
  {Petreczky}}, \bibinfo {author} {\bibfnamefont {Hans-Peter}\ \bibnamefont
  {Schadler}}, \ and\ \bibinfo {author} {\bibfnamefont {Sayantan}\ \bibnamefont
  {Sharma}},\ }\bibfield  {title} {\enquote {\bibinfo {title} {{The topological
  susceptibility in finite temperature QCD and axion cosmology}},}\ }\href
  {\doibase 10.1016/j.physletb.2016.09.063} {\bibfield  {journal} {\bibinfo
  {journal} {Phys. Lett.}\ }\textbf {\bibinfo {volume} {B762}},\ \bibinfo
  {pages} {498--505} (\bibinfo {year} {2016})},\ \Eprint
  {http://arxiv.org/abs/1606.03145} {arXiv:1606.03145 [hep-lat]} \BibitemShut
  {NoStop}%
\bibitem [{\citenamefont {Borsanyi}\ \emph {et~al.}(2016)\citenamefont
  {Borsanyi} \emph {et~al.}}]{javier}%
  \BibitemOpen
  \bibfield  {author} {\bibinfo {author} {\bibfnamefont {Sz.}\ \bibnamefont
  {Borsanyi}} \emph {et~al.},\ }\bibfield  {title} {\enquote {\bibinfo {title}
  {{Calculation of the axion mass based on high-temperature lattice quantum
  chromodynamics}},}\ }\href {\doibase 10.1038/nature20115} {\bibfield
  {journal} {\bibinfo  {journal} {Nature}\ }\textbf {\bibinfo {volume} {539}},\
  \bibinfo {pages} {69--71} (\bibinfo {year} {2016})},\ \Eprint
  {http://arxiv.org/abs/1606.07494} {arXiv:1606.07494 [hep-lat]} \BibitemShut
  {NoStop}%
\bibitem [{\citenamefont {Azcoiti}(2016)}]{vicente}%
  \BibitemOpen
  \bibfield  {author} {\bibinfo {author} {\bibfnamefont {Vicente}\ \bibnamefont
  {Azcoiti}},\ }\bibfield  {title} {\enquote {\bibinfo {title} {{Topology in
  the SU(Nf) chiral symmetry restored phase of unquenched QCD and axion
  cosmology}},}\ }\href {\doibase 10.1103/PhysRevD.94.094505} {\bibfield
  {journal} {\bibinfo  {journal} {Phys. Rev.}\ }\textbf {\bibinfo {volume}
  {D94}},\ \bibinfo {pages} {094505} (\bibinfo {year} {2016})},\ \Eprint
  {http://arxiv.org/abs/1609.01230} {arXiv:1609.01230 [hep-lat]} \BibitemShut
  {NoStop}%
\bibitem [{\citenamefont {Bietenholz}\ \emph {et~al.}(2016)\citenamefont
  {Bietenholz}, \citenamefont {Cichy}, \citenamefont {de~Forcrand},
  \citenamefont {Dromard},\ and\ \citenamefont {Gerber}}]{Wolfgang}%
  \BibitemOpen
  \bibfield  {author} {\bibinfo {author} {\bibfnamefont {Wolfgang}\
  \bibnamefont {Bietenholz}}, \bibinfo {author} {\bibfnamefont {Krzysztof}\
  \bibnamefont {Cichy}}, \bibinfo {author} {\bibfnamefont {Philippe}\
  \bibnamefont {de~Forcrand}}, \bibinfo {author} {\bibfnamefont {Arthur}\
  \bibnamefont {Dromard}}, \ and\ \bibinfo {author} {\bibfnamefont {Urs}\
  \bibnamefont {Gerber}},\ }\bibfield  {title} {\enquote {\bibinfo {title}
  {{The Slab Method to Measure the Topological Susceptibility}},}\ }\bibfield
  {booktitle} {\emph {\bibinfo {booktitle} {{Proceedings, 34th International
  Symposium on Lattice Field Theory (Lattice 2016): Southampton, UK, July
  24-30, 2016}}},\ }\href@noop {} {\bibfield  {journal} {\bibinfo  {journal}
  {PoS}\ }\textbf {\bibinfo {volume} {LATTICE2016}},\ \bibinfo {pages} {321}
  (\bibinfo {year} {2016})},\ \Eprint {http://arxiv.org/abs/1610.00685}
  {arXiv:1610.00685 [hep-lat]} \BibitemShut {NoStop}%
\bibitem [{\citenamefont {Azcoiti}(2017)}]{vicente2}%
  \BibitemOpen
  \bibfield  {author} {\bibinfo {author} {\bibfnamefont {Vicente}\ \bibnamefont
  {Azcoiti}},\ }\bibfield  {title} {\enquote {\bibinfo {title} {{Topology in
  the SU(N$_f$) chiral symmetry restored phase of unquenched QCD and axion
  cosmology. II.}}}\ }\href {\doibase 10.1103/PhysRevD.96.014505} {\bibfield
  {journal} {\bibinfo  {journal} {Phys. Rev.}\ }\textbf {\bibinfo {volume}
  {D96}},\ \bibinfo {pages} {014505} (\bibinfo {year} {2017})},\ \Eprint
  {http://arxiv.org/abs/1704.04906} {arXiv:1704.04906 [hep-lat]} \BibitemShut
  {NoStop}%
\bibitem [{\citenamefont {Azcoiti}\ \emph {et~al.}(2002)\citenamefont
  {Azcoiti}, \citenamefont {Di~Carlo}, \citenamefont {Galante},\ and\
  \citenamefont {Laliena}}]{m1}%
  \BibitemOpen
  \bibfield  {author} {\bibinfo {author} {\bibfnamefont {Vicente}\ \bibnamefont
  {Azcoiti}}, \bibinfo {author} {\bibfnamefont {Giuseppe}\ \bibnamefont
  {Di~Carlo}}, \bibinfo {author} {\bibfnamefont {Angelo}\ \bibnamefont
  {Galante}}, \ and\ \bibinfo {author} {\bibfnamefont {Victor}\ \bibnamefont
  {Laliena}},\ }\bibfield  {title} {\enquote {\bibinfo {title} {{New proposal
  for numerical simulations of thetavacuum - like systems}},}\ }\href {\doibase
  10.1103/PhysRevLett.89.141601} {\bibfield  {journal} {\bibinfo  {journal}
  {Phys. Rev. Lett.}\ }\textbf {\bibinfo {volume} {89}},\ \bibinfo {pages}
  {141601} (\bibinfo {year} {2002})},\ \Eprint
  {http://arxiv.org/abs/hep-lat/0203017} {arXiv:hep-lat/0203017 [hep-lat]}
  \BibitemShut {NoStop}%
\bibitem [{\citenamefont {Azcoiti}\ \emph {et~al.}(2003)\citenamefont
  {Azcoiti}, \citenamefont {Di~Carlo}, \citenamefont {Galante},\ and\
  \citenamefont {Laliena}}]{Azco03}%
  \BibitemOpen
  \bibfield  {author} {\bibinfo {author} {\bibfnamefont {V.}~\bibnamefont
  {Azcoiti}}, \bibinfo {author} {\bibfnamefont {G.}~\bibnamefont {Di~Carlo}},
  \bibinfo {author} {\bibfnamefont {A.}~\bibnamefont {Galante}}, \ and\
  \bibinfo {author} {\bibfnamefont {V.}~\bibnamefont {Laliena}},\ }\bibfield
  {title} {\enquote {\bibinfo {title} {{theta vacuum systems via real action
  simulations}},}\ }\href {\doibase 10.1016/S0370-2693(03)00601-4} {\bibfield
  {journal} {\bibinfo  {journal} {Phys. Lett.}\ }\textbf {\bibinfo {volume}
  {B563}},\ \bibinfo {pages} {117} (\bibinfo {year} {2003})},\ \Eprint
  {http://arxiv.org/abs/hep-lat/0305005} {arXiv:hep-lat/0305005 [hep-lat]}
  \BibitemShut {NoStop}%
\bibitem [{\citenamefont {Azcoiti}\ \emph {et~al.}(2004)\citenamefont
  {Azcoiti}, \citenamefont {Di~Carlo}, \citenamefont {Galante},\ and\
  \citenamefont {Laliena}}]{adgl}%
  \BibitemOpen
  \bibfield  {author} {\bibinfo {author} {\bibfnamefont {V.}~\bibnamefont
  {Azcoiti}}, \bibinfo {author} {\bibfnamefont {G.}~\bibnamefont {Di~Carlo}},
  \bibinfo {author} {\bibfnamefont {A.}~\bibnamefont {Galante}}, \ and\
  \bibinfo {author} {\bibfnamefont {V.}~\bibnamefont {Laliena}},\ }\bibfield
  {title} {\enquote {\bibinfo {title} {{theta dependence of CP**9 model}},}\
  }\href {\doibase 10.1103/PhysRevD.69.056006} {\bibfield  {journal} {\bibinfo
  {journal} {Phys. Rev.}\ }\textbf {\bibinfo {volume} {D69}},\ \bibinfo {pages}
  {056006} (\bibinfo {year} {2004})},\ \Eprint
  {http://arxiv.org/abs/hep-lat/0305022} {arXiv:hep-lat/0305022 [hep-lat]}
  \BibitemShut {NoStop}%
\bibitem [{\citenamefont {Azcoiti}\ \emph {et~al.}(2007)\citenamefont
  {Azcoiti}, \citenamefont {Di~Carlo},\ and\ \citenamefont {Galante}}]{adg}%
  \BibitemOpen
  \bibfield  {author} {\bibinfo {author} {\bibfnamefont {Vecente}\ \bibnamefont
  {Azcoiti}}, \bibinfo {author} {\bibfnamefont {Giuseppe}\ \bibnamefont
  {Di~Carlo}}, \ and\ \bibinfo {author} {\bibfnamefont {Angelo}\ \bibnamefont
  {Galante}},\ }\bibfield  {title} {\enquote {\bibinfo {title} {{Critical
  Behaviour of CP**1 at theta=pi, Haldane's Conjecture, and the Relevant
  Universality Class}},}\ }\href {\doibase 10.1103/PhysRevLett.98.257203}
  {\bibfield  {journal} {\bibinfo  {journal} {Phys. Rev. Lett.}\ }\textbf
  {\bibinfo {volume} {98}},\ \bibinfo {pages} {257203} (\bibinfo {year}
  {2007})},\ \Eprint {http://arxiv.org/abs/0710.1507} {arXiv:0710.1507
  [hep-lat]} \BibitemShut {NoStop}%
\bibitem [{\citenamefont {Azcoiti}\ \emph {et~al.}(2011)\citenamefont
  {Azcoiti}, \citenamefont {Follana},\ and\ \citenamefont {Vaquero}}]{afv}%
  \BibitemOpen
  \bibfield  {author} {\bibinfo {author} {\bibfnamefont {Vicente}\ \bibnamefont
  {Azcoiti}}, \bibinfo {author} {\bibfnamefont {Eduardo}\ \bibnamefont
  {Follana}}, \ and\ \bibinfo {author} {\bibfnamefont {Alejandro}\ \bibnamefont
  {Vaquero}},\ }\bibfield  {title} {\enquote {\bibinfo {title} {{Progress in
  numerical simulations of systems with a $\theta-$vacuum like term: The two
  and three-dimensional Ising model within an imaginary magnetic field}},}\
  }\href {\doibase 10.1016/j.nuclphysb.2011.05.023} {\bibfield  {journal}
  {\bibinfo  {journal} {Nucl. Phys.}\ }\textbf {\bibinfo {volume} {B851}},\
  \bibinfo {pages} {420--442} (\bibinfo {year} {2011})},\ \Eprint
  {http://arxiv.org/abs/1105.1020} {arXiv:1105.1020 [hep-lat]} \BibitemShut
  {NoStop}%
\bibitem [{\citenamefont {Azcoiti}\ \emph {et~al.}(2012)\citenamefont
  {Azcoiti}, \citenamefont {Di~Carlo}, \citenamefont {Follana},\ and\
  \citenamefont {Giordano}}]{adfg}%
  \BibitemOpen
  \bibfield  {author} {\bibinfo {author} {\bibfnamefont {Vicente}\ \bibnamefont
  {Azcoiti}}, \bibinfo {author} {\bibfnamefont {Giuseppe}\ \bibnamefont
  {Di~Carlo}}, \bibinfo {author} {\bibfnamefont {Eduardo}\ \bibnamefont
  {Follana}}, \ and\ \bibinfo {author} {\bibfnamefont {Matteo}\ \bibnamefont
  {Giordano}},\ }\bibfield  {title} {\enquote {\bibinfo {title} {{Critical
  behaviour of the O(3) nonlinear sigma model with topological term at theta=pi
  from numerical simulations}},}\ }\href {\doibase 10.1103/PhysRevD.86.096009}
  {\bibfield  {journal} {\bibinfo  {journal} {Phys. Rev.}\ }\textbf {\bibinfo
  {volume} {D86}},\ \bibinfo {pages} {096009} (\bibinfo {year} {2012})},\
  \Eprint {http://arxiv.org/abs/1207.4905} {arXiv:1207.4905 [hep-lat]}
  \BibitemShut {NoStop}%
\bibitem [{\citenamefont {Azcoiti}\ \emph {et~al.}(2014)\citenamefont
  {Azcoiti}, \citenamefont {Cortese}, \citenamefont {Follana},\ and\
  \citenamefont {Giordano}}]{acfg}%
  \BibitemOpen
  \bibfield  {author} {\bibinfo {author} {\bibfnamefont {V.}~\bibnamefont
  {Azcoiti}}, \bibinfo {author} {\bibfnamefont {G.}~\bibnamefont {Cortese}},
  \bibinfo {author} {\bibfnamefont {E.}~\bibnamefont {Follana}}, \ and\
  \bibinfo {author} {\bibfnamefont {M.}~\bibnamefont {Giordano}},\ }\bibfield
  {title} {\enquote {\bibinfo {title} {{A geometric Monte Carlo algorithm for
  the antiferromagnetic Ising model with ''topological'' term at $\Theta =
  \pi$}},}\ }\href {\doibase 10.1016/j.nuclphysb.2014.04.004} {\bibfield
  {journal} {\bibinfo  {journal} {Nucl. Phys.}\ }\textbf {\bibinfo {volume}
  {B883}},\ \bibinfo {pages} {656--684} (\bibinfo {year} {2014})},\ \Eprint
  {http://arxiv.org/abs/1312.6416} {arXiv:1312.6416 [hep-lat]} \BibitemShut
  {NoStop}%
\bibitem [{\citenamefont {Casher}\ \emph {et~al.}(1974)\citenamefont {Casher},
  \citenamefont {Kogut},\ and\ \citenamefont {Susskind}}]{kogut1}%
  \BibitemOpen
  \bibfield  {author} {\bibinfo {author} {\bibfnamefont {A.}~\bibnamefont
  {Casher}}, \bibinfo {author} {\bibfnamefont {John~B.}\ \bibnamefont {Kogut}},
  \ and\ \bibinfo {author} {\bibfnamefont {Leonard}\ \bibnamefont {Susskind}},\
  }\bibfield  {title} {\enquote {\bibinfo {title} {{Vacuum polarization and the
  absence of free quarks}},}\ }\href {\doibase 10.1103/PhysRevD.10.732}
  {\bibfield  {journal} {\bibinfo  {journal} {Phys. Rev.}\ }\textbf {\bibinfo
  {volume} {D10}},\ \bibinfo {pages} {732--745} (\bibinfo {year}
  {1974})}\BibitemShut {NoStop}%
\bibitem [{\citenamefont {Kogut}\ and\ \citenamefont
  {Susskind}(1975)}]{kogut2}%
  \BibitemOpen
  \bibfield  {author} {\bibinfo {author} {\bibfnamefont {John~B.}\ \bibnamefont
  {Kogut}}\ and\ \bibinfo {author} {\bibfnamefont {Leonard}\ \bibnamefont
  {Susskind}},\ }\bibfield  {title} {\enquote {\bibinfo {title} {{How to Solve
  the eta --> 3 pi Problem by Seizing the Vacuum}},}\ }\href {\doibase
  10.1103/PhysRevD.11.3594} {\bibfield  {journal} {\bibinfo  {journal} {Phys.
  Rev.}\ }\textbf {\bibinfo {volume} {D11}},\ \bibinfo {pages} {3594} (\bibinfo
  {year} {1975})}\BibitemShut {NoStop}%
\bibitem [{\citenamefont {Hamer}\ \emph {et~al.}(1982)\citenamefont {Hamer},
  \citenamefont {Kogut}, \citenamefont {Crewther},\ and\ \citenamefont
  {Mazzolini}}]{kogut3}%
  \BibitemOpen
  \bibfield  {author} {\bibinfo {author} {\bibfnamefont {C.~J.}\ \bibnamefont
  {Hamer}}, \bibinfo {author} {\bibfnamefont {John~B.}\ \bibnamefont {Kogut}},
  \bibinfo {author} {\bibfnamefont {D.~P.}\ \bibnamefont {Crewther}}, \ and\
  \bibinfo {author} {\bibfnamefont {M.~M.}\ \bibnamefont {Mazzolini}},\
  }\bibfield  {title} {\enquote {\bibinfo {title} {{The Massive Schwinger Model
  on a Lattice: Background Field, Chiral Symmetry and the String Tension}},}\
  }\href {\doibase 10.1016/0550-3213(82)90229-2} {\bibfield  {journal}
  {\bibinfo  {journal} {Nucl. Phys.}\ }\textbf {\bibinfo {volume} {B208}},\
  \bibinfo {pages} {413--438} (\bibinfo {year} {1982})}\BibitemShut {NoStop}%
\bibitem [{\citenamefont {Byrnes}\ \emph {et~al.}(2002)\citenamefont {Byrnes},
  \citenamefont {Sriganesh}, \citenamefont {Bursill},\ and\ \citenamefont
  {Hamer}}]{bis}%
  \BibitemOpen
  \bibfield  {author} {\bibinfo {author} {\bibfnamefont {T.}~\bibnamefont
  {Byrnes}}, \bibinfo {author} {\bibfnamefont {P.}~\bibnamefont {Sriganesh}},
  \bibinfo {author} {\bibfnamefont {R.~J.}\ \bibnamefont {Bursill}}, \ and\
  \bibinfo {author} {\bibfnamefont {C.~J.}\ \bibnamefont {Hamer}},\ }\bibfield
  {title} {\enquote {\bibinfo {title} {{Density matrix renormalization group
  approach to the massive Schwinger model}},}\ }\href {\doibase
  10.1103/PhysRevD.66.013002} {\bibfield  {journal} {\bibinfo  {journal} {Phys.
  Rev.}\ }\textbf {\bibinfo {volume} {D66}},\ \bibinfo {pages} {013002}
  (\bibinfo {year} {2002})},\ \Eprint {http://arxiv.org/abs/hep-lat/0202014}
  {arXiv:hep-lat/0202014 [hep-lat]} \BibitemShut {NoStop}%
\bibitem [{\citenamefont {Shimizu}\ and\ \citenamefont
  {Kuramashi}(2014)}]{shimizu}%
  \BibitemOpen
  \bibfield  {author} {\bibinfo {author} {\bibfnamefont {Yuya}\ \bibnamefont
  {Shimizu}}\ and\ \bibinfo {author} {\bibfnamefont {Yoshinobu}\ \bibnamefont
  {Kuramashi}},\ }\bibfield  {title} {\enquote {\bibinfo {title} {{Critical
  behavior of the lattice Schwinger model with a topological term at
  $\theta=\pi$ using the Grassmann tensor renormalization group}},}\ }\href
  {\doibase 10.1103/PhysRevD.90.074503} {\bibfield  {journal} {\bibinfo
  {journal} {Phys. Rev.}\ }\textbf {\bibinfo {volume} {D90}},\ \bibinfo {pages}
  {074503} (\bibinfo {year} {2014})},\ \Eprint {http://arxiv.org/abs/1408.0897}
  {arXiv:1408.0897 [hep-lat]} \BibitemShut {NoStop}%
\bibitem [{\citenamefont {Schwinger}(1962)}]{Schwinger1962}%
  \BibitemOpen
  \bibfield  {author} {\bibinfo {author} {\bibfnamefont {Julian}\ \bibnamefont
  {Schwinger}},\ }\bibfield  {title} {\enquote {\bibinfo {title} {Gauge
  invariance and mass. ii},}\ }\href {\doibase 10.1103/PhysRev.128.2425}
  {\bibfield  {journal} {\bibinfo  {journal} {Phys. Rev.}\ }\textbf {\bibinfo
  {volume} {128}},\ \bibinfo {pages} {2425--2429} (\bibinfo {year}
  {1962})}\BibitemShut {NoStop}%
\bibitem [{\citenamefont {Wiese}(1989)}]{puregauge}%
  \BibitemOpen
  \bibfield  {author} {\bibinfo {author} {\bibfnamefont {U.~J.}\ \bibnamefont
  {Wiese}},\ }\bibfield  {title} {\enquote {\bibinfo {title} {{Numerical
  Simulation of Lattice $\theta$ Vacua: The 2-$d$ U(1) Gauge Theory as a Test
  Case}},}\ }\href {\doibase 10.1016/0550-3213(89)90051-5} {\bibfield
  {journal} {\bibinfo  {journal} {Nucl. Phys.}\ }\textbf {\bibinfo {volume}
  {B318}},\ \bibinfo {pages} {153--175} (\bibinfo {year} {1989})}\BibitemShut
  {NoStop}%
\bibitem [{\citenamefont {Leutwyler}\ and\ \citenamefont
  {Smilga}(1992)}]{smilga}%
  \BibitemOpen
  \bibfield  {author} {\bibinfo {author} {\bibfnamefont {H.}~\bibnamefont
  {Leutwyler}}\ and\ \bibinfo {author} {\bibfnamefont {Andrei~V.}\ \bibnamefont
  {Smilga}},\ }\bibfield  {title} {\enquote {\bibinfo {title} {{Spectrum of
  Dirac operator and role of winding number in QCD}},}\ }\href {\doibase
  10.1103/PhysRevD.46.5607} {\bibfield  {journal} {\bibinfo  {journal} {Phys.
  Rev.}\ }\textbf {\bibinfo {volume} {D46}},\ \bibinfo {pages} {5607--5632}
  (\bibinfo {year} {1992})}\BibitemShut {NoStop}%
\bibitem [{\citenamefont {Coleman}(1976)}]{coleman}%
  \BibitemOpen
  \bibfield  {author} {\bibinfo {author} {\bibfnamefont {Sidney~R.}\
  \bibnamefont {Coleman}},\ }\bibfield  {title} {\enquote {\bibinfo {title}
  {{More About the Massive Schwinger Model}},}\ }\href {\doibase
  10.1016/0003-4916(76)90280-3} {\bibfield  {journal} {\bibinfo  {journal}
  {Annals Phys.}\ }\textbf {\bibinfo {volume} {101}},\ \bibinfo {pages} {239}
  (\bibinfo {year} {1976})}\BibitemShut {NoStop}%
\bibitem [{\citenamefont {Göschl}\ \emph {et~al.}(2017)\citenamefont
  {Göschl}, \citenamefont {Gattringer}, \citenamefont {Lehmann},\ and\
  \citenamefont {Weis}}]{Gattringer}%
  \BibitemOpen
  \bibfield  {author} {\bibinfo {author} {\bibfnamefont {Daniel}\ \bibnamefont
  {Göschl}}, \bibinfo {author} {\bibfnamefont {Christof}\ \bibnamefont
  {Gattringer}}, \bibinfo {author} {\bibfnamefont {Alexander}\ \bibnamefont
  {Lehmann}}, \ and\ \bibinfo {author} {\bibfnamefont {Christoph}\ \bibnamefont
  {Weis}},\ }\bibfield  {title} {\enquote {\bibinfo {title} {{Simulation
  strategies for the massless lattice Schwinger model in the dual
  formulation}},}\ }\href {\doibase 10.1016/j.nuclphysb.2017.09.006} {\bibfield
   {journal} {\bibinfo  {journal} {Nucl. Phys.}\ }\textbf {\bibinfo {volume}
  {B924}},\ \bibinfo {pages} {63} (\bibinfo {year} {2017})},\ \Eprint
  {http://arxiv.org/abs/1708.00649} {arXiv:1708.00649 [hep-lat]} \BibitemShut
  {NoStop}%
\bibitem [{\citenamefont {Azcoiti}\ \emph {et~al.}(1990)\citenamefont
  {Azcoiti}, \citenamefont {di~Carlo},\ and\ \citenamefont {Grillo}}]{mfa}%
  \BibitemOpen
  \bibfield  {author} {\bibinfo {author} {\bibfnamefont {V.}~\bibnamefont
  {Azcoiti}}, \bibinfo {author} {\bibfnamefont {G.}~\bibnamefont {di~Carlo}}, \
  and\ \bibinfo {author} {\bibfnamefont {A.~F.}\ \bibnamefont {Grillo}},\
  }\bibfield  {title} {\enquote {\bibinfo {title} {{A New proposal for
  including dynamical fermions in lattice gauge theories: The Compact QED
  case}},}\ }\href {\doibase 10.1103/PhysRevLett.65.2239} {\bibfield  {journal}
  {\bibinfo  {journal} {Phys. Rev. Lett.}\ }\textbf {\bibinfo {volume} {65}},\
  \bibinfo {pages} {2239} (\bibinfo {year} {1990})}\BibitemShut {NoStop}%
\bibitem [{\citenamefont {Azcoiti}\ \emph {et~al.}(1994)\citenamefont
  {Azcoiti}, \citenamefont {Di~Carlo}, \citenamefont {Galante}, \citenamefont
  {Grillo},\ and\ \citenamefont {Laliena}}]{sch_noi}%
  \BibitemOpen
  \bibfield  {author} {\bibinfo {author} {\bibfnamefont {V.}~\bibnamefont
  {Azcoiti}}, \bibinfo {author} {\bibfnamefont {G.}~\bibnamefont {Di~Carlo}},
  \bibinfo {author} {\bibfnamefont {A.}~\bibnamefont {Galante}}, \bibinfo
  {author} {\bibfnamefont {A.~F.}\ \bibnamefont {Grillo}}, \ and\ \bibinfo
  {author} {\bibfnamefont {V.}~\bibnamefont {Laliena}},\ }\bibfield  {title}
  {\enquote {\bibinfo {title} {{The Schwinger model on the lattice in the
  microcanonical fermionic average approach}},}\ }\href {\doibase
  10.1103/PhysRevD.50.6994} {\bibfield  {journal} {\bibinfo  {journal} {Phys.
  Rev.}\ }\textbf {\bibinfo {volume} {D50}},\ \bibinfo {pages} {6994} (\bibinfo
  {year} {1994})},\ \Eprint {http://arxiv.org/abs/hep-lat/9401032}
  {arXiv:hep-lat/9401032 [hep-lat]} \BibitemShut {NoStop}%
\bibitem [{\citenamefont {D\"urr}(2012)}]{Durr1}%
  \BibitemOpen
  \bibfield  {author} {\bibinfo {author} {\bibfnamefont {Stephan}\ \bibnamefont
  {D\"urr}},\ }\bibfield  {title} {\enquote {\bibinfo {title} {Physics of
  ${\ensuremath{\eta}}^{\ensuremath{'}}$ with rooted staggered quarks},}\
  }\href {\doibase 10.1103/PhysRevD.85.114503} {\bibfield  {journal} {\bibinfo
  {journal} {Phys. Rev. D}\ }\textbf {\bibinfo {volume} {85}},\ \bibinfo
  {pages} {114503} (\bibinfo {year} {2012})}\BibitemShut {NoStop}%
\bibitem [{\citenamefont {D\"urr}\ and\ \citenamefont
  {Hoelbling}(2004)}]{Durr2}%
  \BibitemOpen
  \bibfield  {author} {\bibinfo {author} {\bibfnamefont {Stephan}\ \bibnamefont
  {D\"urr}}\ and\ \bibinfo {author} {\bibfnamefont {Christian}\ \bibnamefont
  {Hoelbling}},\ }\bibfield  {title} {\enquote {\bibinfo {title} {Staggered
  versus overlap fermions: A study in the schwinger model with
  ${N}_{f}=0,1,2$},}\ }\href {\doibase 10.1103/PhysRevD.69.034503} {\bibfield
  {journal} {\bibinfo  {journal} {Phys. Rev. D}\ }\textbf {\bibinfo {volume}
  {69}},\ \bibinfo {pages} {034503} (\bibinfo {year} {2004})}\BibitemShut
  {NoStop}%
\end{thebibliography}
%

\end{document}